\documentclass[aps,pra,twocolumn,groupedaddress,showpacs]{revtex4}

\usepackage{graphicx}
\usepackage{amsmath}
\usepackage{bm}

\begin{document}

%\preprint{}

\title{Spin textures in rotating two-component Bose-Einstein condensates}


%\author{Kenichi Kasamatsu$^1$}
%\author{Makoto Tsubota$^1$}
%\author{Masahito Ueda$^2$}
%\thanks{}

%\affiliation{
%$^1$Department of Physics,
%Osaka City University, Sumiyoshi-Ku, Osaka 558-8585, Japan 
%\\
%$^2$Department of Physics, Tokyo Institute of Technology,  
%Meguro-ku, Tokyo 152-8551, Japan}
\author{Kenichi Kasamatsu}
\affiliation{Department of General Education, 
Ishikawa National College of Technology, Tsubata, Ishikawa 929-0392, Japan}
\author{Makoto Tsubota}
\affiliation{
Department of Physics,
Osaka City University, Sumiyoshi-Ku, Osaka 558-8585, Japan }
\author{Masahito Ueda}
\affiliation{Department of Physics, Tokyo Institute of Technology,  
Meguro-ku, Tokyo 152-8551, Japan}


\date{\today}

\begin{abstract}
We investigate two kinds of coreless vortices with axisymmetric and nonaxisymmetric configurations in rotating two-component Bose-Einstein condensates. Starting from the Gross-Pitaevskii energy functional in a rotating frame, we derive a nonlinear sigma model generalized to the two-component condensates. In terms of a pseudospin representation, an axisymmetric vortex and a nonaxisymmetric one correspond to spin textures referred to as a ``skyrmion" and a ``meron-pair", respectively. A variational method is used to investigate the dependence of the sizes of the stable spin textures on system parameters, and the optimized variational function is found to reproduce well the numerical solution. In the SU(2) symmetric case, the optimal skyrmion and meron-pair are degenerate and transform to each other by a rotation of the pseudospin. An external rf-field that couples coherently the hyperfine states of two components breaks the degeneracy in favor of the meron-pair texture due to an effective transverse pseudomagnetic field. The difference between the intracomponent and intercomponent interactions yields a longitudinal pseudomagnetic field and a ferromagnetic or an antiferromagnetic pseudospin interaction, leading to a meron-pair texture with an anisotropic distribution of vorticity.
\end{abstract}

%\keywords{}

\maketitle

\section{INTRODUCTION}
Since the experimental realization of quantized vortices in alkali atomic Bose-Einstein condensates (BECs) \cite{Matthews,Madison,Haljan,Leanhardt}, there have been a growing interest in new phenomena related to vortices in rotating BECs. A strongly correlated quantum Hall-like phase may appear in systems which rotate so rapidly that the size of the vortex cores becomes comparable with the inter-vortex separation \cite{Baym}. Another direction that has not yet been explored so much concerns rich vortex phases in multicomponent BECs \cite{Ho,Ohmi,Williams,VLeonhardt,Ripoll,Yip,Ripoll3,Mueller,Kasamatsupre,Kawaja,Ruostekoski,Mueller2,Kasamatsu,Kita,Schweikhard}. Since alkali atoms have hyperfine spin, multicomponent BECs can be realized if more than one hyperfine-spin state is populated in the same trap \cite{Hall,Stenger}. The systems described by the multicomponent order parameters allow the excitation of exotic topological defects that have no analogue in systems with a single-component order parameter. For example, while a quantized vortex in a single-component order parameter should have a singular core, it is possible to excite a ``coreless" (non-singular) vortex in multicomponent systems. On the analogy of such topological defects found in other physical systems such as superfluid $^{3}$He \cite{Volovik}, unconventional superconductors \cite{supercond}, quantum Hall systems \cite{Girvin}, nonlinear optics \cite{Chen}, nuclear physics \cite{Skyrmi} and cosmology \cite{cosmo}, creating exotic topological defects in atomic BECs could give us a further insight into related problems and offer a good opportunity to study their physical properties in much greater detail.

This paper addresses the structure and the energetic stability of two kinds of coreless vortices in two-component BECs based on a {\it generalized nonlinear sigma model} (NL$\sigma$M). One of them has an axisymmetric structure, in which the core of one circulating component is filled with the other nonrotating component. This vortex state was created by Matthews {\it et al.} \cite{Matthews}, where they utilized a phase imprinting technique by controlling interconversion between two components spatially and temporally with an external coupling field \cite{Williams}. The structure \cite{Ho} and dynamical stability \cite{Ripoll} of this vortex state have been studied. Here, the spinor nature of the order parameter allows us to interpret the vortex state of the two-component BECs in terms of a ``pseudospin". The pseudospin representation of the two-component BECs reveals that the vortex state observed in Ref. \cite{Matthews} corresponds to a spin texture referred to as a ``skyrmion" \cite{Mueller,Skyrmi} or an ``Anderson-Toulouse vortex" \cite{VLeonhardt,Anderson}. A spin-1 analogue of this texture was created by Leanhardt {\it et al.} \cite{Leanhardt} and theoretically studied in Ref. \cite{Yip,Mizushima,Martikainen}. 

Another coreless vortex which we discuss in this paper has a nonaxisymmetric structure, which may be regarded as a pair of ``merons" \cite{Kasamatsupre,Volovik,Girvin} or ``Mermin-Ho vortices" \cite{Mermin}. This configuration can be realized when each component has one off-centered vortex. Our previous study showed that an external driving field that couples coherently the internal hyperfine states of the two components stabilizes a nonaxisymmetric vortex state \cite{Kasamatsupre}. Then, the internal coupling induces an attractive interaction between the vortex in one component and that in the other, forming a ``vortex-antivortex molecule", which is bound by a domain wall (branch cut) in the relative phase space \cite{Son}. In contrast to a vortex-antivortex pair in conventional superfluid systems, this pair has circulations of the same sign in each individual phase space, but has the opposite sign in the {\it relative phase space}. When more than one vortex molecule is present, one component accomodates vortices and the other accomodates antivortices. 

It is known that the system of two-component BECs such as those studied by a JILA group \cite{Matthews,Hall} approximately possesses the SU(2) symmetry owing to the near-equal scattering lengths within and between the $|F=1,m_{F}=-1\rangle$ and  $|F=2,m_{F}=1\rangle$ hyperfine states of $^{87}$Rb atoms. Here, the two kinds of coreless vortex states are degenerate in a completely SU(2)-symmetric system. However, adding an external perturbation that breaks the SU(2) symmetry would make the stability problem of those vortex states nontrivial. We investigate their structure and stability by exploring the NL$\sigma$M of the two-component BECs, which is derived from a pseudospin representation of the Gross-Pitaevskii energy functional. Using appropriate trial functions for the skyrmion or the meron-pair, we analytically obtain an almost exact spin profile for each case, which leads to a great improvement upon the Thomas-Fermi approximation \cite{Ho}. While our previous paper \cite{Kasamatsupre} used a solution of a classical NL$\sigma$M directly as a variational function, a more general ansatz used here gives accurate profiles of the vortex states. The SU(2) symmetry is broken by the difference between the intracomponent and intercomponent two-body interactions, and by an external field which couples coherently the internal hyperfine states of the two components \cite{Matthews,Matthews2}. Under the pseudospin picture, they give rise to pseudomagnetic fields and a pseudospin-pseudospin interaction, either of which has a great influence on the structure of spin textures. We do not discuss complicated skyrmion excitations with topologically nontrivial spin profiles such as those studied in Ref. \cite{Kawaja,Ruostekoski}.

After formulating our problem in Sec. \ref{formulation}, we derive in Sec. \ref{nonsigma} the NL$\sigma$M  that describes the two-component BECs. Based on the NL$\sigma$M, we determine the optimized structure of an axisymmetric skyrmion using a variational method in Sec. \ref{skyrmion}. We determine the optimized structure of a nonaxisymmetric meron-pair in Sec. \ref{meron}. In Sec. \ref{discuss}, we address the effect of axisymmetry-breaking contributions on the two kinds of spin textures. We conclude this paper in Sec. \ref{concle}. 

\section{FORMULATION OF THE PROBLEM}\label{formulation}
We consider two-component BECs that are condensed into two different hyperfine states $| 1 \rangle$ and $| 2 \rangle$ such as those of $^{87}$Rb atoms. The two-component BECs are assumed to be trapped in the same harmonic potential $V({\bf r})=m(\omega^{2} r^{2} + \omega_{z}^{2} z^{2})/2$. The potential is assumed to rotate at a rotation frequency $\Omega$ about the $z$ axis. Furthermore, their internal states are coupled coherently by an external driving field \cite{Matthews2}. Viewed from the frame of reference corotating with the trap potential, the Gross-Pitaevskii energy functional of our problem reads
\begin{eqnarray}
E[\Psi_{1},\Psi_{2}]=\int d {\bf r} \biggl\{  \sum_{i} \biggl[ \frac{\hbar^{2}}{2m} \biggl| \biggl(\frac{\nabla}{i} - m {\bf \Omega} \times {\bf r} \biggr) \Psi_{i}({\bf r}) \biggr|^{2}  \nonumber \\
+ \biggl( V({\bf r}) - \frac{m}{2} \Omega^{2} r^{2} \biggr) |\Psi_{i}({\bf r})|^{2}  + \frac{g_{i}}{2} |\Psi_{i}({\bf r})|^{4}  
\biggr]  \nonumber \\ 
+ g_{12} |\Psi_{1}({\bf r})|^{2} |\Psi_{2}({\bf r})|^{2} \nonumber \\
 -\hbar \omega_{\rm R} \left[ \Psi_{2}^{\ast}({\bf r}) \Psi_{1}({\bf r}) e^{-i\Delta t} + \Psi_{1}^{\ast}({\bf r}) \Psi_{2}({\bf r}) e^{i\Delta t} \right] \biggr\} \label{INtjos},
\end{eqnarray}
where $\bm{\Omega} = \Omega \hat{\bf z}$, $\Psi_{1}$ and $\Psi_{2}$ denote the condensate wave functions in the two hyperfine states, and $g_{1}$, $g_{2}$, and $g_{12}$ characterize the atom-atom interactions. Here $g_{1}$, $g_{2}$, and $g_{12}$ are expressed in terms of the s-wave scattering lengths $a_{1}$ and $a_{2}$ between atoms in the same hyperfine states and $a_{12}$ between atoms in different hyperfine states as 
\begin{equation}
g_{i}=\frac{4 \pi \hbar^{2} a_{i}}{m} \hspace{4mm} (i=1,2), \hspace{4mm} g_{12}=\frac{4 \pi \hbar^{2} a_{12}}{m}.
\end{equation}
The last two terms in Eq. (\ref{INtjos}) describe a coherent coupling induced by an external driving field, which allows atoms to change their internal state coherently \cite{Matthews2}. Since the driving field is time-dependent, we have introduced the frame in which the driving field is time-independent (i.e., the frame of a laser field). Here, $\omega_{\rm R}(>0)$ is the Rabi frequency and $\Delta$ is a detuning parameter between the external field and the atomic transition. Throughout this paper, we set $\Delta=0$ for simplicity by assuming a complete resonance \cite{tyuu7}. 

It is convenient to measure the length, time and energy scale in units of $b_{\rm ho}=\sqrt{\hbar/m \omega}$, $\omega^{-1}$, and $\hbar \omega$, respectively. Renormalizing the wave function as $\Psi_{i} \rightarrow \sqrt{N} \Psi_{i}/b_{\rm ho}^{3/2}$ with the total particle number $N=N_{1}+N_{2}$, and the energy as $E/\hbar \omega N \rightarrow E$, we obtain 
\begin{eqnarray}
E[\Psi_{1},\Psi_{2}]=\int d {\bf r} \biggl\{ \sum_{i} \biggl[ \frac{1}{2} \biggl| \biggl( \frac{\nabla}{i} - {\bf \Omega} \times {\bf r} \biggr) \Psi_{i} \biggr|^{2} + \tilde{V} |\Psi_{i}|^{2} \nonumber \\
+ \frac{u_{i}}{2} |\Psi_{i}|^{4} \biggr]  + u_{12} |\Psi_{1}|^{2} |\Psi_{2}|^{2} - \omega_{\rm R} \{\Psi_{2}^{\ast} \Psi_{1}  + \Psi_{1}^{\ast} \Psi_{2} \} \biggr\}.
\label{couplendimGP} 
\end{eqnarray}
Here, we denote the trapping potential as $\tilde{V}=\{ (1-\Omega^{2}) r^{2} + \alpha^{2} z^{2} \}/2$ with $\alpha=\omega_{z}/\omega$, the coupling constants as $u_{i}=4 \pi a_{i} N / b_{\rm ho}$ and $u_{12}=4 \pi a_{12} N / b_{\rm ho} $. Since the particles of one component can convert into the other if the internal coherent coupling is present, the total particle number $N = N_{1}+N_{2}$ is conserved, and the normalization of the wave functions can be taken as $\int d {\bf r} ( |\Psi_{1}|^{2} + |\Psi_{2}|^{2} ) = 1$. Minimizing Eq. (\ref{couplendimGP}) with respect to $\Psi_{1}$ and $\Psi_{2}$, we obtain time-independent coupled Gross-Pitaevskii equations
\begin{subequations}
\begin{eqnarray}
\frac{1}{2} \biggl( \frac{\nabla}{i} -{\bf \Omega} \times {\bf r} \biggr)^{2} \Psi_{1} + \tilde{V} \Psi_{1} + u_{1} |\Psi_{1}|^{2} \Psi_{1} \nonumber \\ 
+ u_{12} |\Psi_{2}|^{2} \Psi_{1} - \omega_{\rm R} \Psi_{2} = \mu \Psi_{1} , \\
\frac{1}{2} \biggl( \frac{\nabla}{i} -{\bf \Omega} \times {\bf r} \biggr)^{2} \Psi_{2} + \tilde{V} \Psi_{2} + u_{2} |\Psi_{2}|^{2} \Psi_{2} \nonumber \\ 
+ u_{12} |\Psi_{1}|^{2} \Psi_{2} - \omega_{\rm R} \Psi_{1} = \mu \Psi_{2} . 
\end{eqnarray} \label{timeindGPeq} \end{subequations}
Here, the chemical potential $\mu$, which is common for both components, is determined by the normalization condition. 

\section{A GENERALIZED NONLINEAR SIGMA MODEL}\label{nonsigma}
The pseudospin representation of the order parameter with internal degrees of freedom is useful to obtain a physical interpretation by mapping the system to a magnetic system. Some physical properties of the double-layer quantum Hall system are well understood by projecting the system into a pseudospin space \cite{Girvin}. Also, the spinor order parameter of two-component BECs allows us to analyze this system as a spin-1/2 BEC \cite{VLeonhardt,Mueller,Kasamatsupre,Kawaja,Ruostekoski,Matthews2}. An exact mathematical correspondence can be established between these two systems. In this section, we derive the pseudospin representation of the energy functional, Eq. (\ref{couplendimGP}), which is the NL$\sigma$M that describes the two-component BECs. We assume that $\Psi_{1}$ ($\Psi_{2}$) corresponds to the up (down) component of the spin-1/2 spinor. The nonzero spin projection on the $x$-$y$ plane implies a relative phase coherence between the up- and down-spin components. A similar NL$\sigma$M was also discussed for a two-component Ginzburg-Landau energy functional \cite{Babaev}.

We introduce a normalized complex-valued spinor $\chi=[\chi_{1}({\bf r}), \chi_{2}({\bf r})]^{T}=[|\chi_{1}|e^{i\theta_{1}}, |\chi_{2}|e^{i\theta_{2}}]^{T}$ and decompose the wave function as $\Psi_{i}=\sqrt{\rho_{\rm T}({\bf r})} \chi_{i}({\bf r})$, where $\rho_{\rm T}$ is the total density and thus the spinor satisfies
\begin{equation}
|\chi_{1}|^{2}+|\chi_{2}|^{2}=1.
\label{normalspinor}
\end{equation}
Substitution of the decomposed wave function into Eq. (\ref{couplendimGP}) yields 
\begin{eqnarray}
E = \int d {\bf r} \biggl\{ \frac{1}{2}(\nabla \sqrt{\rho_{\rm T}})^{2} + \frac{\rho_{\rm T}}{2} [ |\nabla \chi_{1}|^{2} + |\nabla \chi_{2}|^{2} \nonumber \\ 
- 2 ({\bf \Omega} \times {\bf r}) (|\chi_{1}|^{2} \nabla \theta_{1} + |\chi_{2}|^{2} \nabla \theta_{2} ) + ({\bf \Omega} \times {\bf r})^{2} ]  \nonumber \\ 
+  \tilde{V} \rho_{\rm T}- 2 \omega_{\rm R} \rho_{\rm T} |\chi_{1}| |\chi_{2}| \cos(\theta_{1}-\theta_{2}) \nonumber \\
+ \frac{1}{2} \rho_{\rm T}^{2} \left[ c_{0} + c_{1} (|\chi_{1}|^{2}-|\chi_{2}|^{2}) 
+c_{2} (|\chi_{1}|^{2} - |\chi_{2}|^{2})^{2}  \right]  \biggr\} \label{spinorene},
\end{eqnarray}
where the new coupling constants are defined as
\begin{subequations}
\begin{eqnarray}
c_{0} \equiv \frac{u_{1}+u_{2}+2u_{12}}{4},  \\
c_{1} \equiv \frac{u_{1}-u_{2}}{2},  \\
c_{2} \equiv \frac{u_{1}+u_{2}-2u_{12}}{4}.
\end{eqnarray}\label{coefficientsc012}
\end{subequations}
The pseudospin density is defined as ${\bf S}=\bar{\chi}({\bf r}) \bm{\sigma} \chi({\bf r})$, where $\bm{\sigma}$ is the Pauli matrix. The explicit expressions of ${\bf S}=(S_{x},S_{y},S_{z})$ are given by 
\begin{subequations}
\begin{eqnarray}
S_{x} = (\chi_{1}^{\ast} \chi_{2} + \chi_{2}^{\ast} \chi_{1}) = 2 |\chi_{1}||\chi_{2}| \cos (\theta_{1} - \theta_{2} ), \\
S_{y} = - i (\chi_{1}^{\ast} \chi_{2} - \chi_{2}^{\ast} \chi_{1}) = -2 |\chi_{1}||\chi_{2}| \sin (\theta_{1} - \theta_{2} ), \\
S_{z} = |\chi_{1}|^{2} - |\chi_{2}|^{2}  ,
\end{eqnarray}\label{spincompon}
\end{subequations}
where the modulus of the total spin is $|{\bf S}|=1$. 

Transformation of Eq. (\ref{spinorene}) to the pseudospin representation can be made as follows. On the kinetic-energy term, we have
\begin{eqnarray}
(\nabla {\bf S})^{2} = (\nabla S_{x})^{2} + (\nabla S_{y})^{2} + (\nabla S_{z})^{2} \nonumber \\
=4 (|\nabla \chi_{1}|^{2} + |\nabla \chi_{2}|^{2} ) - 4 (|\chi_{1}|^{2} \nabla \theta_{1} + |\chi_{2}|^{2} \nabla \theta_{2} )^{2}, \nonumber
\end{eqnarray}
and the second term of Eq. (\ref{spinorene}) becomes
\begin{equation}
\frac{\rho_{\rm T}}{2} \biggl[ \frac{(\nabla {\bf S})^{2}}{4} + \left( |\chi_{1}|^{2} \nabla \theta_{1} + |\chi_{2}|^{2} \nabla \theta_{2} - {\bf \Omega} \times {\bf r} \right)^{2} \biggr] .  \nonumber
\end{equation}
Using Eq. (\ref{normalspinor}) and Eqs. (\ref{spincompon}), we define an effective velocity field
\begin{eqnarray}
{\bf v}_{\rm eff} = |\chi_{1}|^{2} \nabla \theta_{1} + |\chi_{2}|^{2} \nabla \theta_{2} \nonumber \\
= \frac{1}{2} \nabla \Theta + \frac{S_{z}}{2 ( S_{x}^{2}+S_{y}^{2} ) } (S_{y} \nabla S_{x} - S_{x} \nabla S_{y}), \label{effectivevelo}
\end{eqnarray} 
which depends on the gradient of the total phase $\Theta=\theta_{1}+\theta_{2}$ and that of the pseudospin. Thus, we obtain 
\begin{eqnarray}
E = \int  d {\bf r}  \biggl[ \frac{1}{2} (\nabla \sqrt{\rho_{\rm T}})^{2}+ \frac{\rho_{\rm T}}{8} (\nabla {\bf S})^{2} + \frac{\rho_{\rm T}}{2} ( {\bf v}_{\rm eff} - {\bf \Omega} \times {\bf r} )^{2} \nonumber \\
+ \tilde{V} \rho_{\rm T}  - \omega_{\rm R} \rho_{\rm T} S_{x} + \frac{\rho_{\rm T}^{2}}{2} ( c_{0} + c_{1} S_{z} + c_{2} S_{z}^{2}) \biggr]. \nonumber \\ 
\label{nonsigmamod}
\end{eqnarray}
This form is analogous to the classical NL$\sigma$M  for Heisenberg ferromagnets in which only the $(\nabla {\bf S})^{2}$ term appears \cite{Rejan}. In Eq. (\ref{nonsigmamod}), the four degrees of freedom of the original condensate wave functions $\Psi_{1}$ and $\Psi_{2}$ (their amplitudes and phases) are expressed in terms of the total density $\rho_{\rm T}$, the total phase $\Theta$, and two of the spin density components ${\bf S}=(S_{x},S_{y},S_{z})$ (one of them is fixed because of $|{\bf S}|=1$) which are related to the relative density and the relative phase. 

Unique features of Eq. (\ref{nonsigmamod}) that are absent in the classical NL$\sigma$M, are: (i) the total density $\rho_{\rm T}$, which is a prefactor of the $(\nabla {\bf S})^{2}$ term and gives the pseudospin stiffness, is position-dependent because of the trapping potential, (ii) the total phase $\Theta$ appears in the energy functional of Eq. (\ref{nonsigmamod}), (iii) the third term of Eq. (\ref{nonsigmamod}) gives the hydrodynamic kinetic energy $\rho_{\rm T} ({\bf v}_{\rm eff} - {\bf \Omega} \times {\bf r})^{2}/2$ associated with the topological excitation, and (iv) there are several anisotropic terms that break the SU(2) symmetry. The coherent coupling term with the Rabi frequency $\omega_{\rm R}$ works as a transverse (pseudo)magnetic field that aligns the spin along the $x$-axis. The interaction terms including the coefficients $c_{1}$ and $c_{2}$ also break the SU(2) symmetry. The coefficient $c_{1}$ can also be interpreted as a longitudinal (pseudo)magnetic field that aligns the spin along the $z$-axis. The term involving the coefficient $c_{2}$ determines the spin-spin interaction associated with $S_{z}$; it is antiferromagnetic for $c_{2}>0$ and ferromagnetic for $c_{2}<0$ \cite{Kasamatsu}.

\section{AXISYMMETRIC SPIN TEXTURE: A SKYRMION} \label{skyrmion}
Based on the NL$\sigma$M described by Eq. (\ref{nonsigmamod}), we discuss the structure of vortex states in the two-component BECs. In the following sections, we consider the two-dimensional problem by assuming $\omega_{z} \gg \omega$. Then, separating the degrees of freedom of the original wave function as $\Psi_{i}({\bf r})=\psi_{i}(x,y)\phi(z)$, we obtain the dimensionless two-dimensional GP equations
\begin{subequations}
\begin{eqnarray}
\biggl( -\frac{\nabla^{2}}{2} + \frac{r^{2}}{2} + u_{1} |\psi_{1}|^{2} + u_{12} |\psi_{2}|^{2} - \Omega L_{z} \biggr) \psi_{1} = \mu_{1} \psi_{1} ,  \\
\biggl( -\frac{\nabla^{2}}{2} + \frac{r^{2}}{2} + u_{2} |\psi_{2}|^{2} + u_{12} |\psi_{1}|^{2} - \Omega L_{z} \biggr)  \psi_{2} = \mu_{2} \psi_{2}. 
\end{eqnarray} \label{nondimgpeq}
\end{subequations}
Here, we define effective two-dimensional coupling constants $u_{i}=4 \pi a_{i} \eta N$ and $u_{12}=4 \pi a_{12} \eta N$ with $\eta=\int dz |\phi(z)|^{4} / \int dz |\phi(z)|^{2}$. 

\subsection{Numerical results}
\begin{figure}[btp]
\includegraphics[height=0.135\textheight]{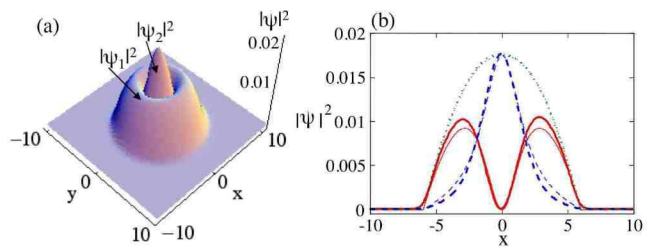}
\caption{(Color online) (a) The density profile of the coreless vortex state consisting of the rotating $\psi_{1}$ component and the nonrotating $\psi_{2}$ component for $u_{1}=u_{2}=u_{12}=1000$ ($c_{0}=1000$, $c_{1}=c_{2}=0$), $\omega_{\rm R}=0$ and $\Omega=0.15$. In this calculation, we fix the total particle number $N=N_{1} + N_{2}$, but do not fix each particle number $N_{i}$. Then, the solution converges to $N_{1}/N_{2}=2.465$. (b) The cross sections of $|\psi_{1}|^{2}$ (solid-curve), $|\psi_{2}|^{2}$ (dashed-curve) and the total density $\rho_{\rm T}$ (dotted-curve) along the $x$-axis at $y=0$, where bold and thin curves represent the results obtained from the numerical calculation and the variational calculation, respectively.} 
\label{axisymvor}
\end{figure}

Figure \ref{axisymvor} shows the density profile of the axisymmetric vortex state obtained by numerically solving Eq. (\ref{nondimgpeq}) for $u_{1}=u_{2}=u_{12}=1000$ ($c_{0}=1000$ and $c_{1}=c_{2}=0$), where the system possesses the SU(2) symmetry if $\omega_{\rm R}=0$. Here, we assume that the $\psi_{1}$ component has one singly-quantized vortex at the center of the trap; we do not discuss the case in which the $\psi_{1}$ component has a multiply quantized vortex. Then, the non-rotating $\psi_{2}$ component is located at the vortex core of the $\psi_{1}$ component; the core size is expanded due to the intercomponent repulsive interaction. As a result, the total density has no singularity and the condensates form a coreless vortex. This vortex structure was created by Matthews {\it et al.} \cite{Matthews} by following the phase imprinting method proposed by Williams and Holland \cite{Williams}. 

\begin{figure}[btp]
\includegraphics[height=0.35\textheight]{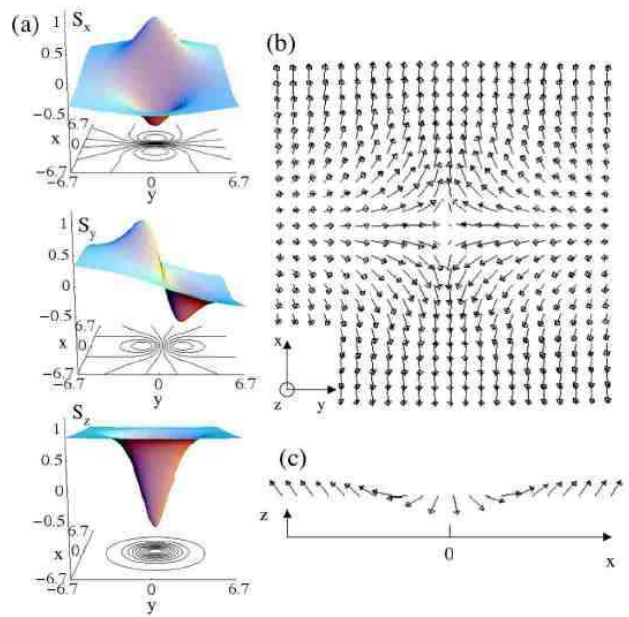}
\caption{(Color online) (a) The pesudospin density ${\bf S}=\bar{\chi} \bm{\sigma} \chi$ corresponding to the vortex state in Fig. \ref{axisymvor}. (b) The vectorial representation of the spin texture projected onto the $x$-$y$ plane in the region $[-6.7 \leq x,y \leq +6.7]$. (c) The cross section of the spin texture along the $x$ axis at $y=0$.} 
\label{skyrmionfig}
\end{figure}
The pseudospin texture corresponding to the axisymmetric vortex state is shown in Fig. \ref{skyrmionfig}, which was also discussed by Leonhardt and Volovik \cite{VLeonhardt} and Mueller \cite{Mueller}. At the center of the cloud, the $\psi_{1}$ component vanishes, and the pseudospin points down in accordance with the definition of the spin $S_{z}$ of Eq. (\ref{spincompon}c). The spin aligns with a hyperbolic distribution with $(S_{x},S_{y}) \propto (x,-y)$ around the singularity at the center [Fig. \ref{skyrmionfig}(b)]. At the edge of the cloud, the $\psi_{2}$ component vanishes, and the pseudospin points up. In between, the pseudospin rolls from down to up continuously as shown in Fig. \ref{skyrmionfig}(c). This cross-disgyration spin texture is often referred to as a ``skyrmion" in analogy to the work of Skyrme \cite{Skyrmi}. 

Axisymmetric spin textures with continuous vorticity was extensively investigated in a field of superfluid $^{3}$He \cite{Volovik}.  When the condensate wave function is parametrized as 
\begin{eqnarray}
\left(
\begin{array}{c}
\psi_{1} \\
\psi_{2}
\end{array}
\right)= 
\sqrt{\rho_{\rm T}} \left(
\begin{array}{c}
e^{i \phi} \cos \frac{\beta(r)}{2} \\
\sin \frac{\beta(r)}{2}
\end{array}
\right), \label{spintexturepara}
\end{eqnarray}
the configuration satisfying the boundary condition $\beta(0)=\pi$ and $\beta(\infty)=0$ is referred to as an ``Anderson-Toulouse (AT)" vortex \cite{VLeonhardt,Anderson}. We also have a ``Mermin-Ho (MH)" vortex or a ``meron" texture with the condition $\beta(0)=\pi$ and $\beta(\infty)=\pi/2$ \cite{VLeonhardt,Mermin}. In the case of superfluid $^{3}$He, a MH vortex is stabilized by the boundary condition imposed by a cylindrical vessel. However, in atomic-BEC system there is no constraint at the boundary; the value $\beta(r)$ at the boundary should be determined self-consistently as discussed later. 

\begin{figure}[btp]
\includegraphics[height=0.18\textheight]{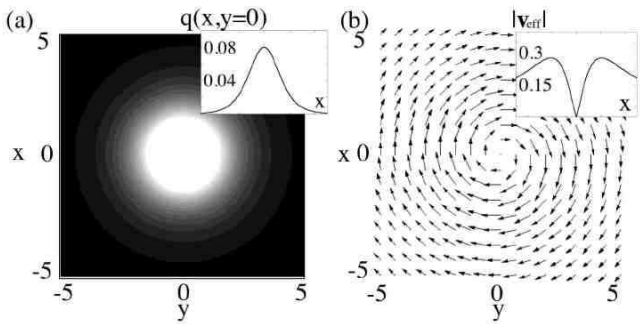}
\caption{(a) The distribution of the topological charge density $q({\bf r})$ of Eq. (\ref{topologicaldensity}) and (b) the vectorial plot of the effective velocity field ${\bf v}_{\rm eff}$ of Eq. (\ref{effectivevelo}), corresponding to the solution shown in Fig. \ref{axisymvor}. The insets represent the cross sections of $q({\bf r})$ and $|{\bf v}_{\rm eff}|$ along the $y=0$ line within the range $-5 \leq x \leq 5$.} 
\label{skyrmtopological}
\end{figure}

It is known that the skyrmion has a topological invariant defined in a two-dimensional system as
\begin{equation}
Q \equiv \frac{1}{8 \pi} \int d {\bf r} \epsilon^{ij} {\bf S} \cdot \partial_{i} {\bf S} \times \partial_{j} {\bf S},
\label{topologicalnumber}
\end{equation}
which is called a topological charge or the Pontryagian index \cite{Girvin}. The skyrmion with any spin profile is shown to have $Q= \pm 1$, whose sign depends on the direction of the circulation of a vortex. The integrand of Eq. (\ref{topologicalnumber}) is the topological charge density associated with the vorticity derived from the effective velocity ${\bf v}_{\rm eff}$ \cite{Mueller,Mermin}: 
\begin{equation}
q({\bf r}) \equiv \frac{1}{8 \pi} \epsilon^{ij} {\bf S} \cdot \partial_{i} {\bf S} \times \partial_{j} {\bf S} = \frac{1}{2\pi} (\nabla \times {\bf v}_{\rm eff} ),
\label{topologicaldensity}
\end{equation}
where we used the relation $\sum_{i} \chi_{i} \nabla \chi_{i}^{\ast} = - \sum_{i} \chi_{i}^{\ast} \nabla \chi_{i}$ $(i=1,2)$ in obtaining the last equality. The topological charge density $q({\bf r})$ characterizes the spatial distribution of the skyrmion. Figure \ref{skyrmtopological} shows the spatial distribution of $q({\bf r})$ and the corresponding ${\bf v}_{\rm eff}$-field. The topological charge is distributed around the center and, contrary to a conventional vortex in a single-component condensate, $|{\bf v}_{\rm eff}|$ vanishes at the center. This makes a coreless vortex without a density dip in the total density. 

\subsection{Variational analysis}
To study the physical properties of the skyrmion in more detail, we make a variational analysis based on the NL$\sigma$M in Eq. (\ref{nonsigmamod}). The original NL$\sigma$M (Eq. (\ref{nonsigmamod}) with only the ($\nabla {\bf S})^{2}$ term) admits a skyrmion solution and explicit analytic expressions are known \cite{Rejan}. Here we take a more general form of the skyrmion solution as a variational function; the skyrmion solution of Fig. \ref{skyrmionfig} may be parametrized as \cite{Sinova}
\begin{eqnarray}
S_{x}=\frac{4 \lambda x e^{-\alpha r^{2}/2}}{r^{2}+4 \lambda^{2} e^{-\alpha r^{2}}}, \nonumber \\ 
S_{y}=\frac{-4 \lambda y e^{-\alpha r^{2}/2}}{r^{2}+4 \lambda^{2} e^{-\alpha r^{2}}}, \label{skyrmansatz} \\ 
S_{z}=\frac{r^{2}-4\lambda^{2} e^{-\alpha r^{2}}}{r^{2}+4\lambda^{2} e^{-\alpha r^{2}}} \nonumber
\end{eqnarray}
with $|{\bf S}|=1$. Equations (\ref{skyrmansatz}) with $\alpha=0$ correspond to the explicit skyrmion solution of the classical NL$\sigma$M \cite{Rejan}. The variational parameters $\lambda$ and $\alpha$ determine the size and the shape of the skyrmion. Typically, $\lambda$ represents the size of the domain in which the spin is reversed and $\alpha$ the degree of spatial variation of the spin inversion. If we take both $\lambda$ and $\alpha$ as variational parameters, the number of particles in each pseudospin component is not conserved (we will discuss below the situation in which each particle number is fixed). Therefore, the energetically stable configuration of the skyrmion brings about an imbalance in the particle number of each component. Such population imbalance can be controlled by the turning-off time of the coupling drive in the phase imprinting method \cite{Matthews,Williams}. For the limit $\lambda \rightarrow 0$ $(\infty)$ with fixed $\alpha$ all spins point up (down) along the $z$-axis, which means perfect polarization of the particle number as $N_{1} \rightarrow N$ $(N_{2} \rightarrow N)$. Alternately, for fixed $\lambda$, the limit $N_{1} \rightarrow N$ $(N_{2} \rightarrow N)$ corresponds to $\alpha \rightarrow + \infty$ $(- \infty)$.  Because the $\psi_{1}$ component has one singly-quantized vortex at the center, the total phase $\Theta$ in Eq. (\ref{nonsigmamod}) is given by 
\begin{equation}
\Theta = \tan^{-1} \frac{y}{x}. \label{skyrmtotpha}
\end{equation}
Substituting Eqs. (\ref{skyrmansatz}) and (\ref{skyrmtotpha}) into Eq. (\ref{nonsigmamod}) yields 
\begin{eqnarray}
E=\int d {\bf r} \biggl\{ \frac{1}{2} (\nabla \sqrt{\rho_{\rm T}})^{2} + V_{\rm eff} \rho_{\rm T} -\omega_{\rm R} \rho_{\rm T} \frac{4 \lambda x e^{-\alpha r^{2}/2} }{r^{2}+4\lambda^{2} e^{-\alpha r^{2}}}  \nonumber \\
+\frac{\rho_{\rm T}^{2}}{2} \biggl[ c_{0} + c_{1} \frac{r^{2}-4\lambda^{2} e^{-\alpha r^{2}}}{r^{2}+4\lambda^{2} e^{-\alpha r^{2}}} + c_{2} \biggl( \frac{r^{2}-4\lambda^{2} e^{-\alpha r^{2}}}{r^{2}+4\lambda^{2} e^{-\alpha r^{2}}} \biggr)^{2} \biggr] \biggr\}. \nonumber \\
\label{symmskyreneg}
\end{eqnarray}
Now the degrees of freedom in the energy functional have been reduced to the total density $\rho_{\rm T}$ and the two variational parameters $\lambda$ and $\alpha$. Here, we have introduced an effective confining potential 
\begin{eqnarray}
V_{\rm eff} \equiv \frac{r^{2}+4 \lambda^{2} e^{-\alpha r^{2}} [ (\alpha r^{2} +1)^{2} +1 ] }{2 (r^{2}+4\lambda^{2} e^{-\alpha r^{2}})^{2}} \nonumber \\ 
- \frac{\Omega r^{2}}{r^{2}+4\lambda^{2} e^{-\alpha r^{2}}} + \frac{r^{2}}{2}, 
\label{skyrmeffpot}
\end{eqnarray} 
which is radially symmetric and determines the shape of the total density $\rho_{\rm T}$. Note that the term with the Rabi frequency $\omega_{\rm R}$ in Eq. (\ref{symmskyreneg}) breaks the axisymmetry of the problem; we will discuss the effect of this term in Sec. \ref{meron} and \ref{discuss}. Ignoring this term, we can calculate the total density $\rho_{\rm T}$ by solving the equation
\begin{eqnarray}
-\frac{(\nabla^{2} \sqrt{\rho_{\rm T}})}{2 \sqrt{\rho_{\rm T}}} + V_{\rm eff} + \rho_{\rm T} \biggl\{ c_{0} +  c_{1} \frac{r^{2}-4\lambda^{2} e^{-\alpha r^{2}}}{r^{2}+4\lambda^{2} e^{-\alpha r^{2}}}  \nonumber \\ 
+ c_{2} \biggl( \frac{r^{2}-4\lambda^{2} e^{-\alpha r^{2}}}{r^{2}+4\lambda^{2} e^{-\alpha r^{2}}} \biggr)^{2} \biggr\} = \mu,
\label{totalden1}
\end{eqnarray} 
where the chemical potential $\mu$ is fixed by the normalization condition $\int d {\bf r} \rho_{\rm T} = 1$. 

\begin{figure}[btp]
\includegraphics[height=0.24\textheight]{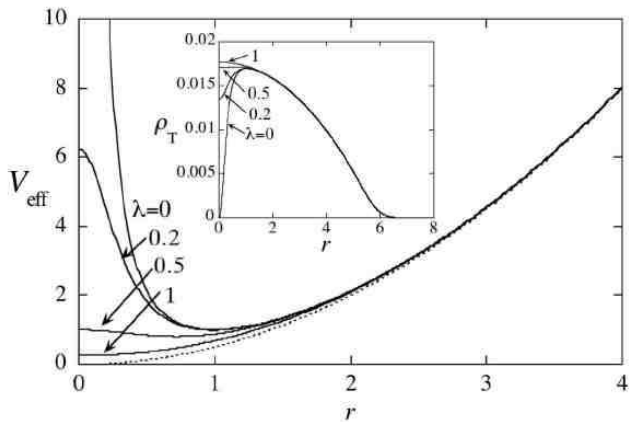}
\caption{(a) The effective potential $V_{\rm eff}$ of Eq. (\ref{skyrmeffpot}) with $\alpha=0$ (for $\Omega=0$ and $\omega_{\rm R}=0$) for several values of $\lambda$. A pure harmonic potential (obtained for $\lambda \rightarrow \infty$) is shown by a dotted curve. The inset shows the corresponding total density profiles $\rho_{\rm T}$ for $c_{0}=1000$.} 
\label{skyrmeffpotfig}
\end{figure}
Before proceeding the analysis further, it is instructive to investigate the form of the effective potential $V_{\rm eff}$. For simplicity, let us first consider the case with $\alpha=0$, i.e., the classical skyrmion solution. Figure \ref{skyrmeffpotfig} shows the effective potential $V_{\rm eff}$ and the total density profile $\rho_{\rm T}$ calculated numerically from Eq. (\ref{totalden1}) for several values of $\lambda$ for $c_{0}=1000$, $c_{1}=c_{2}=0$ and $\Omega=0$. For $\lambda=0$ ($N = N_{1}$), the value of $V_{\rm eff}$ at $r=0$ diverges, which implies that the total density $\rho_{\rm T}$ should vanish at the center. This corresponds to the singular vortex core of the $\psi_{1}$ component. With increasing $\lambda$ the central peak of $V_{\rm eff}$ decreases and the total density becomes smooth because of the appearance of the nonrotating $\psi_{2}$ component, which fills the vortex core of the $\psi_{1}$ component. The limit $\lambda \rightarrow \infty$ $(N \rightarrow N_{2})$ describes the vanishing $\psi_{1}$ component with vorticity. Then, the effective potential becomes a pure harmonic potential and the total density becomes $|\psi_{2}|^{2}$ exactly. This character of $V_{\rm eff}$ also holds for $\alpha \neq 0$. To describe $\rho_{\rm T}$ except for $\lambda \simeq 0$, therefore, we can employ the Thomas-Fermi approximation which neglects the quantum-pressure term $-(\nabla^{2} \sqrt{\rho_{\rm T}})/2 \sqrt{\rho_{\rm T}}$ in Eq. (\ref{symmskyreneg}). Then, the total density is directly given from Eq. (\ref{totalden1}). 

\subsubsection{SU(2)-symmetric condensates}
\begin{figure}[btp]
\includegraphics[height=0.23\textheight]{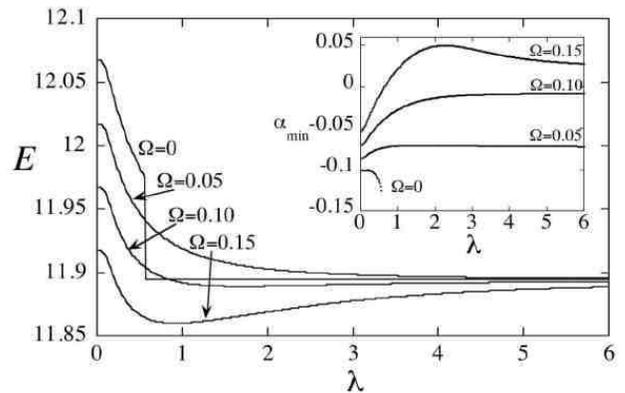}
\caption{The total energy as a function of the skyrmion size $\lambda$ for several values of the rotation frequency $\Omega$ ($\Omega=$0, 0.05, 0.1, 0.15) for $c_{0}=10^{3}$ and $c_{1}=c_{2}=0$. The energy of the nonvortex state is 11.894. The inset shows the optimized value of $\alpha$ as a function of $\lambda$.} 
\label{skyrmeenerot}
\end{figure}
Using the condition $\int d {\bf r} \rho_{\rm T}=1$, we numerically calculate the chemical potential in Eq. (\ref{totalden1}) with the Thomas-Fermi approximation (the region with negative $\rho_{\rm T}$ is neglected) and the total energy $E$ as a function of $\lambda$ and $\alpha$ from the resulting total density $\rho_{\rm T}$. Figure \ref{skyrmeenerot} shows the total energy $E$ of the axisymmetric vortex state as a function of $\lambda$ for $c_{0}=10^{3}$, $c_{1}=c_{2}=0$ and several values of the rotation frequency $\Omega$, where the value of $\alpha$ is optimized for each $\lambda$. For values of $\Omega$ higher than 0.17, more vortices enter the system. In the case without rotation ($\Omega=0$), the energy decreases with $\lambda$ and suddenly drops to the value at which the system has no vorticity, i.e., complete polarization of the particle number as $N=N_{2}$ (see the inset of Fig. \ref{skyrmeenerot}), where there is no energy minimum corresponding to the vortex state for fixed $\lambda$ because $\alpha \rightarrow - \infty$. This means that the global minimum for $\Omega=0$ is a nonvortex state for {\it any values of $N_{1}$ and $N_{2}$ under the fixed total particle number $N$}, because the interaction-energy terms now satisfy the SU(2) symmetry so that the energy of the nonvortex state is degenerate under the change of the ratio $N_{1}/N_{2}$ with fixed $N$. However, beyond a certain critical value of $\Omega$, there appears an energy minimum as shown in Fig. \ref{skyrmeenerot}. Since the minimized energy is lower than that of the nonvortex state at $\lambda \rightarrow +\infty$, the external rotation can make the corresponding skyrmion texture stable globally. 

\begin{figure}[btp]
\includegraphics[height=0.22\textheight]{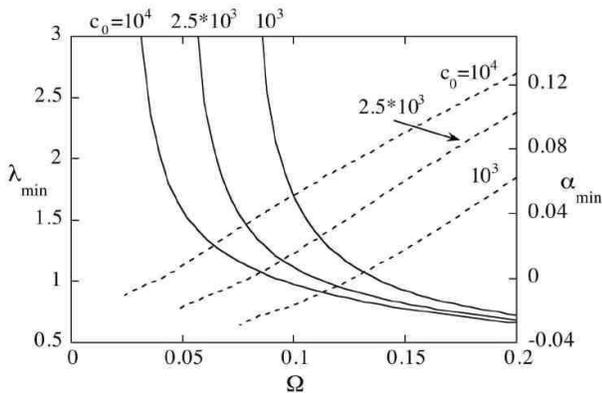}
\caption{The optimized value of the variational parameter $\lambda$ (solid curve) and that of $\alpha$ (dashed curve) as a function of rotation frequency $\Omega$ for $c_{0}=10^{3}$, $2.5 \times 10^{3}$, $10^{4}$ and $c_{1}=c_{2}=0$.} 
\label{optimizedsize}
\end{figure}

Figure \ref{optimizedsize} shows the values of $\lambda$ and $\alpha$ that give the minimum of the total energy for $c_{0}=10^{3}$, $2.5 \times 10^{3}$, and $10^{4}$. The size of the skyrmion decreases with increasing $\Omega$, as revealed by a decrease (an increase) in $\lambda_{\rm min}$ ($\alpha_{\rm min}$). The energy minimum appears at a certain critical frequency of $\Omega$, where $\lambda_{\rm min}$ ($\alpha_{\rm min}$) diverges to $+\infty$ ($-\infty$) and the critical frequency decreases as $c_{0}$ increases. Therefore, the condensates with larger $c_{0}$ can have a stable skyrmion at a lower rotation frequency. However, the full numerical calculations of Eq. (\ref{nondimgpeq}) show that additional vortices are nucleated and form a lattice of skyrmions beyond $\Omega \simeq 0.17$, 0.11, 0.05 for $c_{0}=10^{3}$, $2.5\times 10^{3}$, $10^{4}$, respectively. These structures cannot be described by Eq. (\ref{skyrmansatz}) and the investigation of such periodic solutions remains to be made. From the comparison with the numerical solution as shown in Fig. \ref{axisymvor}(b), the optimized variational functions Eqs. (\ref{skyrmansatz}) and (\ref{skyrmtotpha}) reproduce an almost exact numerical solution. Hence, our approach improves greatly the analytic treatment for the vortex states based on the usual Thomas-Fermi approximation studied in Ref. \cite{Ho}. 

\begin{figure}[btp]
\includegraphics[height=0.22\textheight]{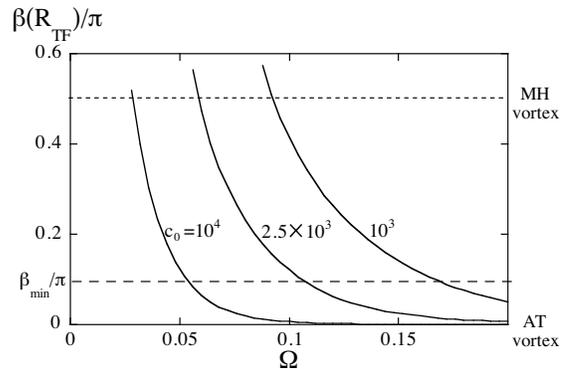}
\caption{The bending angle $\beta(r)=\cos^{-1} S_{z}$ at the Thomas-Fermi boundary $r=R_{\rm TF}$ at which the total density $\rho_{\rm T}$ vanishes, as a function of $\Omega$. The optimized values of the variational parameters $\lambda$ and $\alpha$ in Fig. \ref{optimizedsize} are used.} 
\label{optimizebeta}
\end{figure}

The bending angle $\beta(r)=\cos^{-1} S_{z}$ in Eq. (\ref{spintexturepara}) decreases smoothly from $\pi$ at $r=0$ as $r$ increases. Figure \ref{optimizebeta} shows the values of $\beta(r)$ evaluated at the Thomas-Fermi boundary $r=R_{\rm TF}$ at which the total density $\rho_{\rm T}$ vanishes, as a function of rotation frequency $\Omega$. The value $\beta(R_{\rm TF})$ can change arbitrary with $\Omega$ by varying the ratio $N_{1}/N_{2}$, which implies that an intermediate configuration between a MH vortex ($\beta(R_{\rm TF})=\pi/2$) and an AT vortex ($\beta(R_{\rm TF})=0$) can be made thermodynamically stable. It is the ratio $N_{1}/N_{2}$ that determines $\beta(R_{\rm TF})$, which does not change for fixed $N_{1}/N_{2}$ as explained in Sec. \ref{fixpar}. We also note that there exists a minimum value of $\beta(R_{\rm TF})$ ($\sim$ 0.09), below which additional vortices are nucleated. Therefore, an AT vortex can never be the ground state of the SU(2)-symmetric two-component condensates. A similar discussion is made in Ref. \cite{Mizushima,Martikainen} for the case of ferromagnetic spin-1 BECs.

\subsubsection{$c_{1}$ and $c_{2}$ dependence}
\begin{figure}[btp]
\includegraphics[height=0.22\textheight]{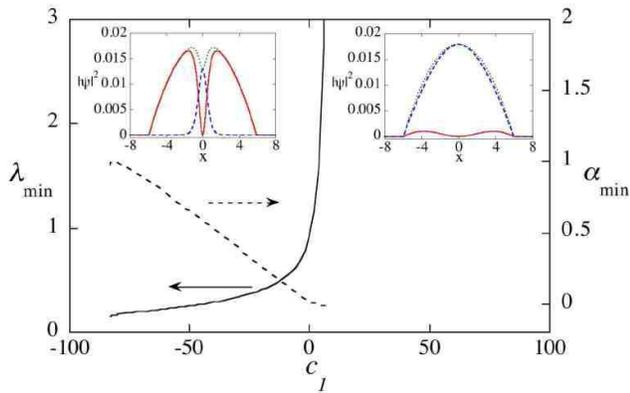}
\caption{(Color online) The optimized values of the variational parameters $\lambda$ (solid curve) and $\alpha$ (dashed curve) as a function of the longitudinal (pseudo)magnetic field $c_{1}$ for $c_{0}=10^{3}$, $c_{2}=0$, and $\Omega=0.15$. Insets show the cross sections of the variational function $|\psi_{1}|^{2}$ (solid curve), $|\psi_{2}|^{2}$ (dashed curve), and the total density $\rho_{\rm T}$ (dotted curve) along the $y=0$ line (left: $c_{1}=-50$, right: $c_{1}=8$).} 
\label{optimc1case}
\end{figure}

Next, we discuss the effects of $c_{1}$ and $c_{2}$ in Eq. (\ref{symmskyreneg}), the values of which are controllable by changing the scattering lengths. The coefficient $c_{1}$, proportional to the difference of the intracomponent interactions $u_{1}$ and $u_{2}$, may be regarded as a (pseudo)magnetic field that aligns the spin along the $z$-axis; the positive (negative) $c_{1}$ aligns the spins downward (upward). Figure \ref{optimc1case} shows the optimized value of $\lambda$ and $\alpha$ as a function of $c_{1}$ for $\Omega=0.15$ and $c_{0}=10^{3}$. For the negative $c_{1}$ side the size of a stable skyrmion shrinks as $|c_{1}|$ increases, which implies the spin alignment to upward with increasing a fraction of the rotating $\psi_{1}$ component. In this case, an AT vortex can be thermodynamically stable. Below $\lambda$ at $c_{1} \simeq -85$, the Thomas-Fermi approximation breaks down for small $\lambda$ owing to the appearance of the singular vortex core in $\rho_{\rm T}$. For the positive $c_{1}$ side the value of $\lambda$ increases rapidly and eventually goes to infinity (concurrently, $\alpha$ goes to $-\infty$), corresponding to the complete ``spin-down" alignment (occupation of the nonrotating $\psi_{2}$ component only). This dependence on the sign of $c_{1}$ reflects the fact that the occupation of the component with vorticity costs a larger energy than that of the nonrotating component. 

\begin{figure}[btp]
\includegraphics[height=0.22\textheight]{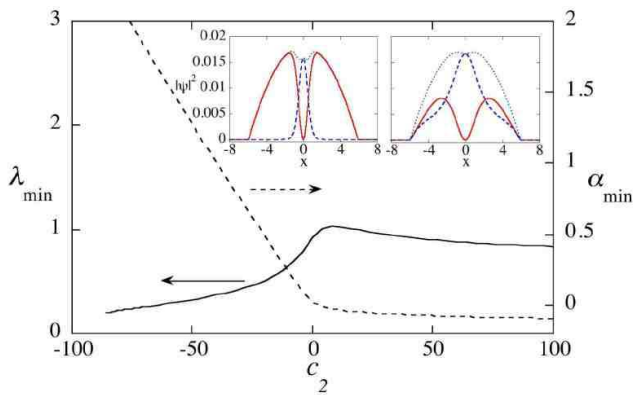}
\caption{(Color online) The optimized values of the variational parameters $\lambda$ (solid curve) and $\alpha$ (dashed curve) as a function of the spin-spin interaction strength $c_{2}$ for $c_{0}=10^{3}$, $c_{1}=0$, and $\Omega=0.15$. The insets show the cross sections of the variational function $|\psi_{1}|^{2}$ (solid curve), $|\psi_{2}|^{2}$ (dashed curve), and the total density $\rho_{\rm T}$ (dotted curve) along the $y=0$ line (left: $c_{2}=-50$, right: $c_{2}=50$).} 
\label{optimc2case}
\end{figure}

The sign of $c_{2}$ determines the nature of the spin-spin interaction associated with $S_{z}$. For positive $c_{2}$ ($u_{1}+u_{2}>2u_{12}$) the interaction is antiferromagnetic, preferring to the spatial mixture of the spin-up and spin-down components. While for negative $c_{2}$ ($u_{1}+u_{2}<2u_{12}$), the system enters the ferromagnetic phase, where the spin domains are spontaneously formed. This feature also appears in the stable structure of the skyrmion. Figure \ref{optimc2case} shows the optimized values of $\lambda$ and $\alpha$ as a function of $c_{2}$ for $\Omega=0.15$, $c_{0}=10^{3}$ and $c_{1}=0$. For the antiferromagnetic case $c_{2}>0$, we see no significant change of the spin profile compared with that of $c_{2}=0$ (the SU(2) symmetric case). For the ferromagnetic case $c_{2}<0$, the variations of $\lambda$ and $\alpha$ are similar to that of Fig \ref{optimc1case}, where the size of the skyrmion shrinks with $|c_{2}|$. Since each particle number is not conserved in this calculation, the energetically favorable configuration in a ferrmagnetic phase tends to an overall spin polarization, i.e., complete polarization of the particle number. In this regime, there are two energy minima, one of which leads to the perfect polarization of the $\psi_{1}$ vortex state (corresponding to the results shown in Fig. \ref{optimc2case}) and the other leads to the $\psi_{2}$ nonvortex state (characterized by $\lambda \rightarrow \infty$ and $\alpha \rightarrow - \infty$; not shown), respectively, and which state possesses the global stability depends on the rotation frequency. 

\subsubsection{Energy minimization under the fixed particle number of each component}\label{fixpar}
So far, we have allowed a change in the particle number of each component to calculate the minimized skyrmion size. However, the experiments at JILA on two-component BECs \cite{Hall} were made under the condition in which each particle number is fixed. This restriction can be taken into account by noticing the relation
\begin{equation}
\frac{N_{1}}{N_{2}} = \frac{\int d {\bf r} \rho_{T} (1 + S_{z})}{\int d {\bf r} \rho_{T} (1 - S_{z})};
\label{ratioparti}
\end{equation}
for a given $\lambda$ the value of $\alpha$ is uniquely determined by Eq. (\ref{ratioparti}), where both $\rho_{\rm T}$ and $S_{z}$ are functions of two variational parameters. Therefore, the energy minimization can be done with respect to one variational parameter which we choose to be $\lambda$. As discussed before, this ratio determines the boundary value of the bending angle $\beta(r)$.

\begin{figure}[btp]
\includegraphics[height=0.44\textheight]{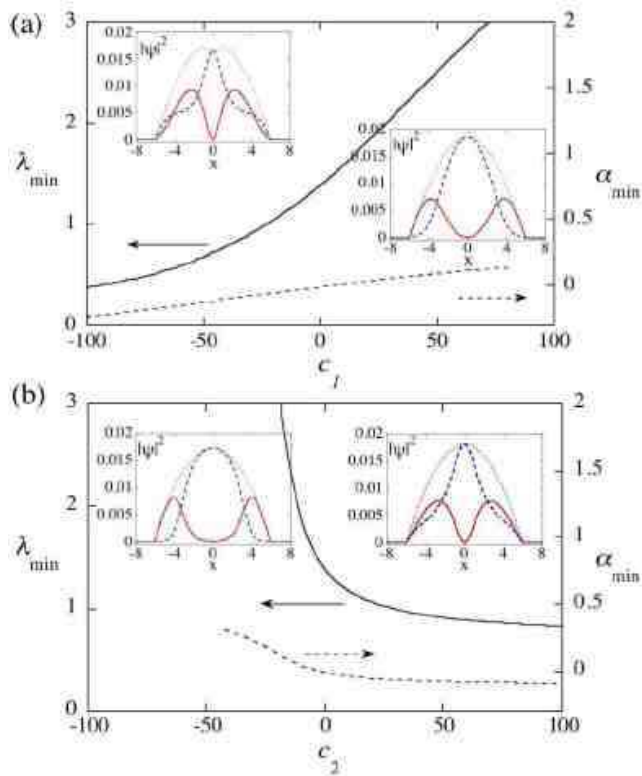}
\caption{(Color online) The optimized values of the variational parameters $\lambda$ (solid curve) and $\alpha$ (dashed curve) as a function of (a) $c_{1}$ and (b) $c_{2}$ under the condition of a fixed particle number of each component $N_{1}/N_{2}=1$ for $c_{0}=10^{3}$, $\Omega=0.15$ ($c_{2}=0$ and $c_{1}=0$ for (a) and (b), respectively). The insets show the cross sections of the variational functions $|\psi_{1}|^{2}$ (solid curve), $|\psi_{2}|^{2}$ (dashed curve), and the total density $\rho_{\rm T}$ (dotted curve) along the $y=0$ line ((a) left: $c_{1}=-50$, right: $c_{1}=50$ and (b) left: $c_{2}=-30$, right: $c_{2}=30$).} 
\label{optimfixNcase}
\end{figure}

Here, we consider the situation of the equal particle number $N_{1}/N_{2}=1$ and investigate the stable size of the skyrmion as done before. In this case, we find that the stable size is not affected by the change of $\Omega$. Figure \ref{optimfixNcase} shows the optimized values of $\lambda$ and $\alpha$ as a function of $c_{1}$ and $c_{2}$. In this case, no complete polarization of the particle number occur, so that the skyrmion may exist for all values of $c_{1}$ and $c_{2}$; in this case, however, even if the optimal variational parameters exist, they do not ensure a local minimum of the total energy. In Fig. \ref{optimfixNcase}(a), for positive $c_{1}$ the divergence of $\lambda_{\rm min}$ seen in Fig. \ref{optimc1case} is suppressed due to the conservation of each particle number. Alternately, to enlarge the ``spin-down" domain where the $\psi_{2}$ component is occupied, the size of the vortex core of the $\psi_{1}$ component becomes large by increasing both $\lambda_{\rm min}$ and $\alpha_{\rm min}$. For negative $c_{1}$, the domain with the rotating $\psi_{1}$ component tends to increase. As a result, the vortex core of the $\psi_{1}$ component shrinks and a part of the $\psi_{2}$ component is pushed out of the $\psi_{1}$ component. Here, the total energy of the skyrmion with the same $|c_{1}|$ is always lower for positive $c_{1}$ than for negative $c_{1}$. This is because it is favorable for the $\psi_{1}$ component with the relation $u_{1}>u_{2}$ to spread spatially by the centrifugal effect with a vortex, and increase the ``spin-down" domain. This is consistent with the experimental observation \cite{Matthews} that the vortex in the component having a larger intracomponent interaction is stable; otherwise it is unstable (see also the argument of the dynamical stability in Ref. \cite{Ripoll}). In the $c_{2}$ dependence, while we see no drastic change in the antiferromagnetic $c_{2}>0$ range, for $c_{2}<0$ a rapid increase of $\lambda_{\rm min}$ and $\alpha_{\rm min}$ with increasing $|c_{2}|$ is seen in Fig. \ref{optimfixNcase}(b). This means that the spin-up or spin-down domains grow and their boundary becomes sharp with increasing $|c_{2}|$. 
 
\section{NONAXISYMMETRIC SPIN TEXTURE: A MERON-PAIR} \label{meron}
\subsection{Numerical results}
\begin{figure}[btp]
\includegraphics[height=0.27\textheight]{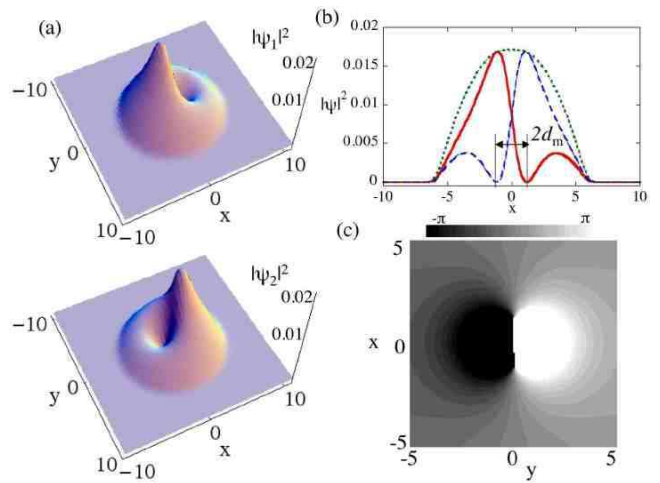}
\caption{(Color online) (a) The density profile of the condensates $|\psi_{1}|^{2}$ and $|\psi_{2}|^{2}$ for $u_{1}=u_{2}=u_{12}=1000$ ($c_{0}=10^{3}$, $c_{1}=c_{2}=0$), $\Omega=0.15$ and $\omega_{\rm R}=0.05$. The two components have the same particle number. (b) The cross sections of $|\psi_{1}|^{2}$ (solid-curve), $|\psi_{2}|^{2}$ (dashed-curve) and the total density $\rho_{\rm T}$ (dotted-curve) along the $x$-axis at $y=0$, where bold and thin curves represent the results obtained from the numerical calculation and those obtained from the variational calculation, respectively. (c) The gray scale plot of the relative phase $\phi=\theta_{1}-\theta_{2}$.} 
\label{nonaxivor}
\end{figure}

In this section, we discuss a nonaxisymmetric vortex state. This structure can be regarded as a spin texture consisting of a pair of merons \cite{Girvin} or MH vortices \cite{Volovik,Mermin}. Our previous study showed that this meron-pair can be stabilized thermodynamically in the presence of rotation and internal coherent coupling \cite{Kasamatsupre}. A typical solution of this structure is shown in Fig. \ref{nonaxivor}. Each component has one off-axis vortex and the density peak of one component is located at the vortex core of the other component. This results in a coreless vortex in which the total density has no singularity. 

Because of the presence of the coherent coupling, the profile of the relative phase $\phi({\bf r})=\theta_{1}-\theta_{2}$ plays an important role in optimizing the structure, which is shown in Fig. \ref{nonaxivor}(c). The relative phase shows that the central region is characterized by the configuration of a vortex-antivortex pair. In other words, the two vortices are connected by a domain wall of the relative phase with the $2 \pi$ difference, which is characteristic of a topological soliton in two-component BECs with the internal coherent coupling. Son and Stephanov \cite{Son} obtained an exact form of the domain wall in a homogeneous one-dimensional system. In our case, the one-dimensional profile of $\phi({\bf r})$ along the $x=0$ line approximately takes a form 
\begin{equation}
\phi(0,y)=-4 \tan^{-1} e^{k y} + C,
\end{equation} 
where $k^{-1}=(|\psi_{1}||\psi_{2}|/2\omega_{\rm R} \rho_{\rm T})^{1/2}$ gives a characteristic size of the domain wall, and the constant $C=2\pi$ ($0$) for $y>0$ ($<0$) makes the branch cut at $y=0$ shown in Fig. \ref{nonaxivor}(c). Then, the vortex in one component and that in the other can be bound by this domain wall, forming a ``vortex molecule" \cite{Kasamatsupre}. 

\begin{figure}[btp]
\includegraphics[height=0.34\textheight]{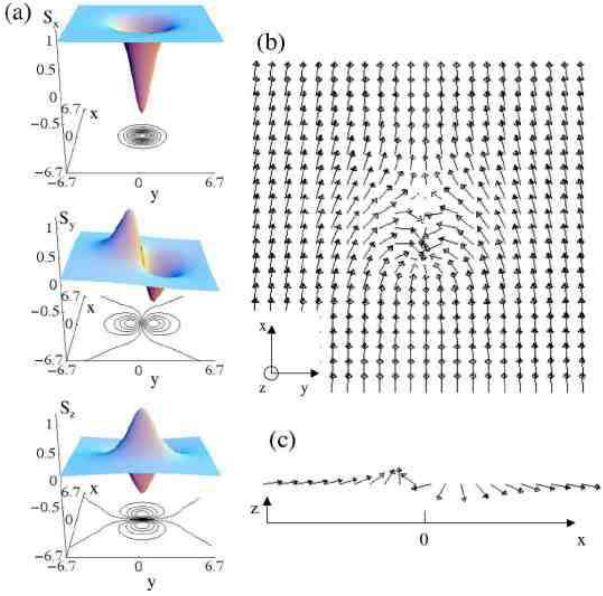}
\caption{(Color online) (a) The pesudospin density ${\bf S}=\bar{\chi} \bm{\sigma} \chi$ corresponding to the vortex state in Fig. \ref{nonaxivor}. (b) The vectorial representation of the spin texture projected onto the $x$-$y$ plane in the region $[-6.7 \leq x,y \leq 6.7]$. (c) The cross section of the spin texture (b) along the $x$ axis at $y=0$.} 
\label{meronfig}
\end{figure}
The pseudospin profiles corresponding to the solution of Fig. \ref{nonaxivor} are shown in Fig. \ref{meronfig}(a). While the profile of $S_{x}$ is axisymmetric, $S_{y}$ and $S_{z}$ show dipole structures. The structure of ${\bf S}$ has a form similar to what is obtained by exchanging $S_{x}$ and $S_{z}$ in Fig. \ref{skyrmionfig}. The corresponding spin texture is shown in Fig. \ref{meronfig}(b) and (c). The spins are oriented in the $x$ direction everywhere except in the central domain-wall region, where they tumble rapidly by $2\pi$. There exist two points corresponding to the locations of vortices at which ${\bf S}$ is parallel to the $z$-axis. The spin around the singularity with ${\bf S} = +{\bf \hat{z}}$ (${\bf S} = -{\bf \hat{z}}$) has a radial (hyperbolic) distribution, characterized by $(S_{x},S_{y}) \propto (-x,-y)$ ($\propto (x,-y)$), and rotates through 90$^{\circ}$ as it goes outward to become perpendicular to the $x$-axis. This texture is known as a ``radial-hyperbolic" pair of merons, which has been discussed in the study of topological defects in superfluid $^{3}$He \cite{Volovik} and a double-layer quantum Hall system \cite{Girvin}. Since the trapping potential is axisymmetric, the energy is degenerate with respect to the orientation of the molecular polarization. If the molecule is polarized along the $y$-axis, the texture forms a ``circular-hyperbolic" pair \cite{Kasamatsupre} and can therefore continuously change into the radial-hyperbolic form. 

We find that, when $u_{1}=u_{2}=u_{12}$ ($c_{1}=c_{2}=0$), the topological charge density $q({\bf r})$ and the effective velocity field ${\bf v}_{\rm eff}$ exhibit a {\it radially isotropic profile} as in Fig. \ref{skyrmtopological}, even though each component forms a nonaxisymmetric vortex configuration. As we discuss later, this is due to the fact that the meron-pair texture and the axisymmetric skyrmion texture are equivalent as a topological excitation in the case of $u_{1}=u_{2}=u_{12}$; they can transform to each other through an overall rotation of the pseudospin. Therefore, the Rabi coupling alone does not break the axisymmetry of the topological excitation. However, inclusion of both the Rabi coupling and unequal coupling constants $u_{1} \neq u_{2} \neq u_{12}$ induces an anisotropy of the meron-pair as discussed in Sec. \ref{discuss}.  

\subsection{Variational analysis}
In the case of $u_{1}=u_{2}=u_{12}$ ($c_{1}=c_{2}=0$), the spin profile of the meron-pair in Fig. \ref{meronfig} is radially isotropic and may be parametrized as \cite{Kasamatsupre}
\begin{eqnarray}
S_{x}=\frac{r^{2}-4\lambda^{2} e^{-\alpha r^{2}}}{r^{2}+4\lambda^{2} e^{-\alpha r^{2}}}, \nonumber \\ 
S_{y}=\frac{-4 \lambda y e^{-\alpha r^{2}/2}}{r^{2}+4 \lambda^{2} e^{-\alpha r^{2}}}, \label{meronprofiloe} \\
S_{z}=\frac{-4 \lambda x e^{-\alpha r^{2}/2}}{r^{2}+4 \lambda^{2} e^{-\alpha r^{2}}}. \nonumber
\end{eqnarray}
Here we assume that the vortex molecule is polarized along the $x$-axis. Compared with the ansatz solution in Eq. (\ref{skyrmansatz}) for a skyrmion, only $S_{x}$ and $S_{z}$ are exchanged. In this case, the ratio of the particle number given in Eq.  (\ref{ratioparti}) is always unity, because $S_{z}$ is an odd function. In Ref. \cite{Kasamatsupre}, we set $\alpha=0$ to simplify the variatioal problem, and obtained a reasonable agreement with the numerical solution. The full variational analysis with two variational parameters $\lambda$ and $\alpha$ gives almost perfect quantitative agreement with the numeircal one as seen below.

The locations of the vortex cores are determined by two extremes of $S_{z}$, given by 
\begin{equation}
x^{2} = 4 \lambda^{2} e^{-\alpha x^{2}}. 
\label{moldistance}
\end{equation}
It solution is represented as $x=\pm W(4 \alpha \lambda^{2})/\sqrt{\alpha}$ with the product log function $W(z)$ \cite{wolfram,reldomain}. Then, it is natural to take the form of the total phase $\Theta$ as
\begin{equation}
\Theta = \tan^{-1} \frac{y}{x-2 \lambda e^{-\alpha r^{2}/2}} + \tan^{-1} \frac{y}{x+2 \lambda e^{-\alpha r^{2}/2}}.
\label{merontotpha}
\end{equation}
Substituting Eq. (\ref{meronprofiloe}) and Eq. (\ref{merontotpha}) into Eq. (\ref{nonsigmamod}), we obtain the total energy similar to Eq. (\ref{symmskyreneg}):  
\begin{eqnarray}
E=\int d {\bf r} \biggl[ \frac{1}{2} (\nabla \sqrt{\rho_{\rm T}})^{2} + V_{\rm eff} \rho_{\rm T} +c_{0} \frac{\rho_{\rm T}^{2}}{2} \nonumber \\
 - \omega_{\rm R} \rho_{T} \frac{r^{2} - 4\lambda^{2} e^{-\alpha r^{2}}}{r^{2}+4\lambda^{2} e^{-\alpha r^{2}}}  \biggr]. \label{meroneneg}
\end{eqnarray}
Here, the effective confining potential $V_{\rm eff}$ is the same as in Eq. (\ref{skyrmeffpot}). Contrary to the axisymmetric skyrmion, the internal coupling term with $\omega_{\rm R}$ does not break the radial symmetry of the problem.

Because Eq. (\ref{meroneneg}) gives the same energy for any value of $\lambda$ and $\alpha$ as that of the axisymmetric skyrmion given by Eq. (\ref{symmskyreneg}), if $\omega_{\rm R}=0$, i.e., for the SU(2) symmetric case, the skyrmion and the meron-pair have the same optimal values of $\lambda$ and $\alpha$ and their energies are degenerate. In other words, the energy is degenerate with respect to the rotation of the overall pseudospin between the two spin textures. We can see that turning on the Rabi term $\omega_{\rm R}$ always decreases the free energy of the meron-pair from any solutions of vortex states with the topological charge $Q=+1$ for $\omega_{\rm R}=0$. Therefore, adding the infinitesimal value of $\omega_{\rm R}$ is enough to stabilize the nonaxisymmetric configuration if $c_{1}=c_{2}=0$. 

\begin{figure}[btp]
\includegraphics[height=0.44\textheight]{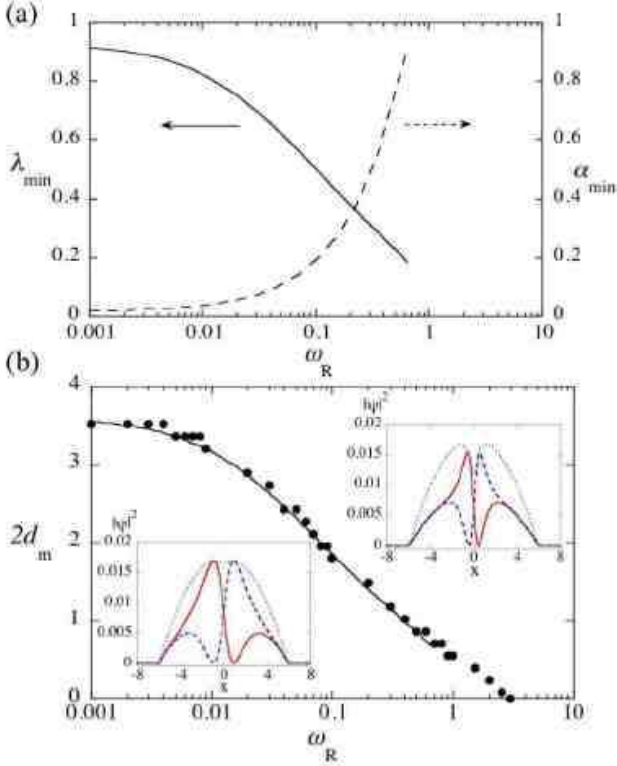}
\caption{(Color online) (a) The optimized values of the variational parameters $\lambda$ (solid curve) and $\alpha$ (dashed curve) as a function of $\omega_{\rm R}$ for $c_{0}=10^{3}$, $\Omega=0.15$, $c_{1}=0$ and $c_{2}=0$. (b) The corresponding size of the meron-pair $2d_{\rm m}$ calculated from Eq. (\ref{moldistance}) (see Fig. \ref{nonaxivor}(b) for definition) for $c_{0}=10^{3}$. We also show $2d_{\rm m}$ obtained from numerical simulations as filled circles. The insets show the cross sections of the variational functions $|\psi_{1}|^{2}$ (solid curve), $|\psi_{2}|^{2}$ (dashed curve), and the total density $\rho_{\rm T}$ (dotted curve) along the $y=0$ line (left: $\omega_{\rm R}=0.1$, right: $\omega_{\rm R}=0.5$).}  
\label{optimRabicase}
\end{figure}

Since the Rabi frequency may be regarded as a transverse magnetic field along the $x$-axis, the stable size of the meron-pair as a function of $\omega_{\rm R}$ shows a similar behavior as that shown in Fig. \ref{optimc1case}. The difference is that the Rabi term is proportional to $\rho_{\rm T}$ instead of $\rho_{\rm T}^{2}$, which changes the behavior from that of Fig. \ref{optimc1case}. In the presence of slow rotation \cite{tyuu2}, we calculate the minimized values of $\lambda$ and $\alpha$ as a function of $\omega_{\rm R}$, the result being shown in Fig. \ref{optimRabicase}(a). As $\omega_{\rm R}$ increases, the minimized value of $\lambda$ ($\alpha$) becomes smaller (larger), and eventually vanishes (diverges). This behavior corresponds to a decrease in the size $d_{\rm m}$ of the meron-pair (see in Fig. \ref{nonaxivor}(b) for the definition) as shown in Fig. \ref{optimRabicase}(b), where $d_{\rm m}$ is calculated from Eq. (\ref{moldistance}). This indicates that the binding of the meron-pair becomes stronger with increasing $\omega_{\rm R}$, which is caused by the tension of a domain wall in the relative phase between the two-component wave functions \cite{Kasamatsupre,Son}. The variational result agrees perfectly with the numerical result. Beyond $\omega_{\rm R} \simeq 3.0$ the separation $d_{\rm m}$ vanishes, where the locations of the density nodes overlap despite the intercomponent repulsive interaction. 

The physical origin of the binding of the meron-pair, i.e., the vortex molecule, was discussed previously \cite{Kasamatsupre}. For well-separated merons, the repulsive interaction between them originates from the second and third terms of Eq. (\ref{nonsigmamod}), which are the gradient energy of the pseudospin and the hydrodynamic kinetic energy of the ${\bf v}_{\rm eff}$-field. On the other hand, the attractive force between two merons are estimated from a tension $T_{\rm d}$ of the domain wall of the relative phase to be $\sim T_{\rm d} d_{\rm m}$; for a homogeneous system $T_{\rm d} = 8 |\psi_{1}|^{2} |\psi_{2}|^{2} k / \rho_{\rm T}$ with the characteristic domain size $k^{-1}= (|\psi_{1}||\psi_{2}|/2\omega_{\rm R} \rho_{\rm T})^{1/2}$ \cite{Son}, we have $T_{\rm d} \propto \sqrt{\omega_{\rm R}}$. Then, the competition between the repulsive force and the attractive force creates an energy minimum so that the two vortices can form a bound pair. The contribution of the other terms are almost constant except for small $d_{\rm m}$ ($<0.30$ for $c_{0}=10^{3}$), where the vortex core appears in the total density. Then, the Thomas-Fermi approximation cannot apply to the evaluation of the total energy; $\lambda_{\rm min}$ drops suddenly to zero with increasing $\omega_{\rm R}$ because of the lacking of the energy barrier associated with the quantum pressure of $\rho_{\rm T}$. The numerical result shows that the separation $d_{\rm m}$ decreases smoothly to zero with $\omega_{\rm R}$.

\subsection{Basis transformation of the pseudospin}
In the SU(2) symmetric case ($c_{1}=c_{2}=0$), a further insight of the nonaxisymmetric vortex is obtained by rotating the basis of the spinor so that the internal coupling becomes simpler. Note that the spin profile of a skyrmion [Fig. \ref{skyrmionfig}] and a meron -pair [Fig. \ref{meronfig}] is connected only through the exchange of $S_{x}$ and $S_{z}$. In this case, the Rabi term makes the $x$-axis as a preferred one.  Actually, if we rotate the spinors as $\psi_{\pm}=(\psi_{1} \pm \psi_{2})/\sqrt{2}$ (the basis along with the $x$-axis), the coupled Gross-Pitaevskii equations (\ref{timeindGPeq}) become 
\begin{subequations}
\begin{eqnarray}
\biggl[ \frac{1}{2} \biggl( \frac{\nabla}{i} -{\bf \Omega} \times {\bf r} \biggr)^{2} + \tilde{V} + c_{0} \rho_{\rm T} \biggr] \psi_{+} 
= (\mu + \omega_{\rm R}) \psi_{+} , \nonumber \\
\biggl[ \frac{1}{2} \biggl( \frac{\nabla}{i} -{\bf \Omega} \times {\bf r} \biggr)^{2} + \tilde{V} + c_{0} \rho_{\rm T} \biggr]  \psi_{-} 
 = (\mu - \omega_{\rm R}) \psi_{-} . \nonumber 
\end{eqnarray} \label{timeindGPeqhenkan} 
\end{subequations}
Here, the internal coupling is just the chemical potential difference between the ``$+$" and ``$-$" components. Then, the nonaxisymmetric structure in Fig. \ref{nonaxivor} is transformed to the axisymmetric vortex state of Fig. \ref{axisymvor}, i.e., a skyrmion. Here, the vortex core of the ``$+$" component is filled with the nonrotating ``$-$" component.  As one increases $\omega_{\rm R}$ the number of the ``$-$" particle drops because of a decrease in the chemical potential of $\psi_{-}$, and the vortex cores eventually become empty, which corresponds to the overlap of the vortex cores.

\section{EFFECTS OF SYMMETRY-BREAKING TERMS}\label{discuss}
In this section, we discuss the effect of the symmetry breaking terms, which are neglected in the last two sections, by numerically solving Eqs. (\ref{nondimgpeq}). For an axisymmetric skyrmion texture, the Rabi term breaks the axisymmetry of the spin texture, and for a meron-pair the $c_{1}$ and $c_{2}$ term cause symmetry breaking. 

\begin{figure}[btp]
\includegraphics[height=0.52\textheight]{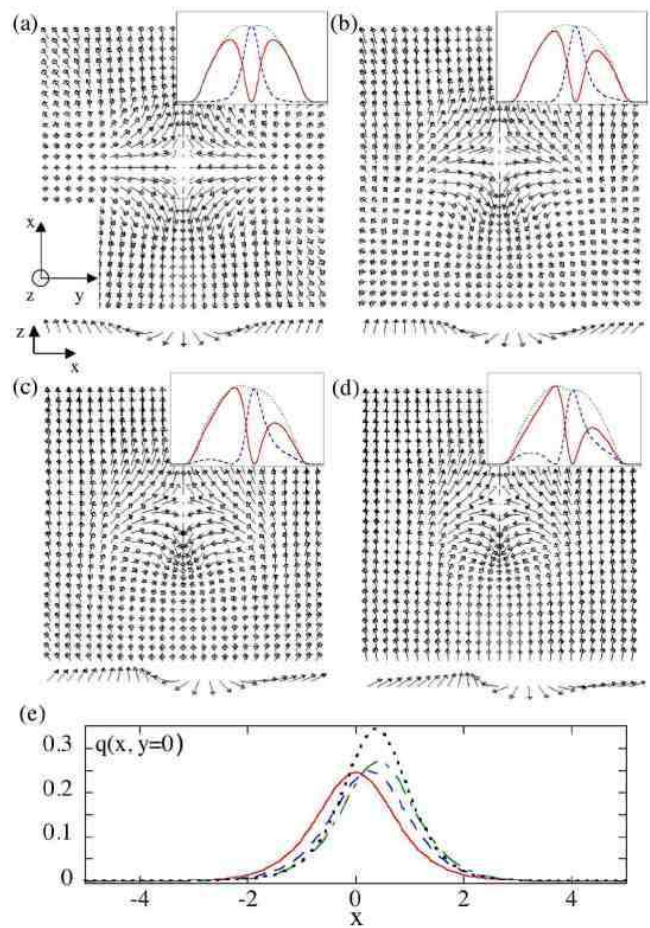}
\caption{(Color online) Equilibrium structures of the spin texture in the presence of a longitudinal magnetic field ($c_{1}$) and a transverse magnetic field ($\omega_{\rm R}$) in the region $[-4 \leq x,y \leq +4]$. The vectorial plots in the $x$-$y$ plane and along the $y$=0 line are shown. The values of the parameters are $\Omega=0.15$, $c_{0}=1000$, $c_{1}=-10$, $c_{2}=0$, and (a) $\omega_{\rm R}=0$, (b) $\omega_{\rm R}=0.02$, (c) $\omega_{\rm R}=0.05$, (a) $\omega_{\rm R}=0.10$. The cross sections of $|\psi_{1}|^{2}$ (solid curve), $|\psi_{2}|^{2}$ (dashed curve), and the total density $\rho_{\rm T}$ (dotted curve) along the $y=0$ line is also shown within the range $[-8 \leq x \leq +8]$. (e) The distributions of the topological charge density $q({\bf r})$ along the $y=0$ line for $\omega_{\rm R} = 0$ (solid), 0.02 (dashed), 0.05 (dashed-dotted), 0.1 (dotted).} 
\label{texturechange}
\end{figure}

First, we neglect the spin-spin interaction term ($c_{2}$-term) and investigate the dependence of the stable structure on $c_{1}$ and $\omega_{\rm R}$, which corresponds to a situation in which a longitudinal and transverse (pseudo)magnetic field are applied simultaneously. We first prepare a stable axisymmetric skyrmion by applying the longitudinal magnetic field $c_{1}=-10$, and then turn on the transverse magnetic field $\omega_{\rm R}$. Figures \ref{texturechange}(a)-(d) show the variation of the spin texture as well as the cross section of the condensate densities along the $y=0$ line when the Rabi frequecy $\omega_{\rm R}$ is increased. The transverse magnetic field shifts the skyrmion to the off-center and convert it into a meron. Correspondingly, another meron enters from outward and forms an {\it asymmetric} vortex molecule shown in Figs. \ref{texturechange}(c) and (d). This change is a second-order transition; there is no energy barrier to destabilize the axisymmetric skyrmion with respect to the transverse magnetic field. Figure \ref{texturechange}(e) shows the topological charge density $q({\bf r})$ along the $x$-axis (the $y=0$ line). Although its distribution is shifted from the center with increasing $\omega_{\rm R}$, there is no dramatic change in the global shape from the isotropic one. As $\omega_{\rm R}$ increases further, the peak of $q({\bf r})$ gets back to the center and its value goes to infinity when the two merons merge.

\begin{figure}[btp]
\includegraphics[height=0.48\textheight]{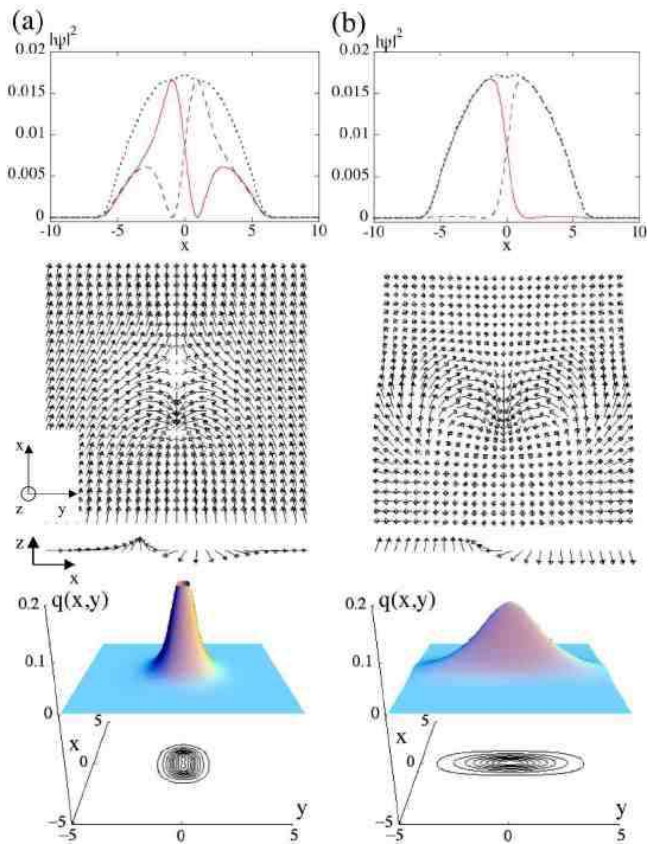}
\caption{(Color online) Equilibrium structures of the spin texture in the presence of a spin-spin interaction $c_{2}$ and a transverse magnetic field ($\omega_{\rm R}$). The parameter values are $\Omega=0.15$, $c_{0}=1000$, $c_{1}=0$, $\omega_{\rm R}=0.05$, and (a) $c_{2}=20$, (b) $c_{2}=-20$. (top) The cross sections of $|\psi_{1}|^{2}$ (solid curve), $|\psi_{2}|^{2}$ (dashed curve), and the total density $\rho_{\rm T}$ (dotted curve) along the $y=0$ line is also shown. (middle) The corresponding spin texture in the region $[-4 \leq x,y \leq +4]$. The vectorial plots in the $x$-$y$ plane and along the $y$=0 line are shown. (bottom) The spatial distributions of the topological charge density $q({\bf r})$.} 
\label{c2ferroantiferro}
\end{figure}

Next, we study the effects of the spin-spin interaction term ($c_{2}$ term) on the stable structure of a meron-pair under a transverse magnetic field with $\omega_{\rm R}=0.05$ ($c_{1}$ is fixed to be zero). Then, $c_{2}$ changes the structure of the meron-pair dramatically, in contrast to the case of $c_{1}$. Figure \ref{c2ferroantiferro} shows the equilibrium structure of the condensate density, the spin texture and the topological charge density $q({\bf r})$ for the antiferromagnetic case ($c_{2}=-20$) and the ferromagnetic case ($c_{2}=20$). For the antiferromagnetic case there is no significant difference in the density and spin profile, compared with the solution of $c_{2}=0$ in Figs. \ref{nonaxivor} and \ref{meronfig}. However, the topological charge density, i.e., vorticity, is distributed anisotropically in such a manner that the distribution is elongated along the direction of polarization of the meron-pair. For the ferromagnetic case, the spin domains are formed, which gives rise to a considerable change in the density and spin profile as seen in Fig. \ref{c2ferroantiferro}(b). Most of the spins align up or down along the $z$-axis on both side, sandwiching the domain wall across which the spin flips rapidly. If the Rabi frequency $\omega_{\rm R}$ is increased further, the spins on both sides are laid along the $x$-axis, converting into the well-defined meron-pair as in Fig. \ref{meronfig}. Then, the topological charge density is distributed in the direction perpendicular to the direction of polarization of the meron-pair, being concentrated on the domain-wall region. This anisotropy of the meron-pair gives an interesting situation when the condensates undergo a rapid rotation; the anisotropic interaction between the meron-pairs generates a distorted lattice of the ``vortex molecules" \cite{Kasamatsupre}, which is an unique feature of the rotating two-component system with the internal coherent coupling. 

\section{CONCLUSION}\label{concle}
In conclusion, we have discussed the coreless vortex states and the corresponding spin textures in rotating two-component BECs with and without internal coherent coupling. The axisymmteric and nonaxisymmetric structure of the spin texture have been discussed by exploring the NL$\sigma$M derived from the Gross-Pitaevskii energy functional. The variational function of the spin profile, which is based on the exact solution of the classical NL$\sigma$M, provides a good description of the coreless vortex states. This variational method also improves greatly the usual analytical approach based on the Thomas-Fermi approximation \cite{Ho}. In the case of SU(2) symmetry, these two spin textures are equivalent topological excitations and transform to each other by a global rotation of the pseudospin. We have discussed the effect of the SU(2)-symmetry breaking contributions, inequality of the three coupling constants $u_{1}$, $u_{2}$ and $u_{12}$ ($c_{1}$ and $c_{2}$ term) and the Rabi frequency $\omega_{\rm R}$, on the spin textures. These contributions are interpreted as a longitudinal and transverse magnetic field and a spin-spin interaction in the NL$\sigma$M. For an axisymmetric skyrmion (a nonaxisymmetric meron-pair), the $c_{1}$ and $c_{2}$ term ($\omega_{\rm R}$ term) affect the optimized size of the topological excitation, but do not break the radial symmetry of the problem. Inclusion of all those terms makes a topological excitation characterized by an anisotropic distribution of the vorticity and the topological charge density. 

One of the open problems is to study vortex states in a ferromagnetic regime. Rich spin textures are expected with the combination of the ferromagnetic feature leading to spin-domain formation and a rotational effect, e.g., a rotating domain-wall cross (``propellers") \cite{Malomed} and ``serpentine" vortex sheets \cite{Kasamatsu}. The problem of their stability remains to be investigated. Furthermore, it is necessary to study in more detail lattices of spin textures which appear in the rapidly rotating two-component BECs, where we can expect a richer phase diagram and structural phase transitions of the vortex phases \cite{Kasamatsupre,Zhai}. It is also of interest to extend the pseudospin model to two-component fermion systems in view of the rapid progress in ultracold fermions after the observation of condensation of fermionic pairs \cite{Regal}; rapidly rotating two-component fermions \cite{Ghosh} may exhibit phenomena analogous to what has been discussed in double-layer quantum Hall systems.

\begin{acknowledgments}
The authors are grateful to M.M. Salomaa and E.J. Mueller for useful comments. K.K. and M.T. acknowledge support by a Grant-in-Aid for Scientific Research (Grants No. 15$\cdot$5955 and No. 15341022) by the JSPS. M.U. acknowledges support by a Grant-in-Aid for Scientific Research (Grant No.15340129) by the MEXT of Japan, and a CREST program by JST. 
\end{acknowledgments}



\begin{thebibliography}{99}
\bibitem{tyuu7}
The detuning parameter $\Delta$ in Eq. (\ref{INtjos}) ignored in this paper can be utilized to control the ratio of the particle number, because it gives the chemical potential difference between two components. As a result, an additional term $\Delta \rho_{\rm T} S_{z} / 2$ appears in the NL$\sigma$M. This term thus acts as a magnetic field that aligns the spin along the $z$-axis just like the $c_{1}$-term, being proportional to $\rho_{\rm T}$ instead of $\rho_{\rm T}^{2}$.
\bibitem{reldomain}
It is easy to see that this variational function also describes the domain-wall structure in the relative phase space. If we take $0<\phi<+\pi$ ($-\pi<\phi<0$) in the region $y>0$ ($y<0$), the profile of $\phi=\tan^{-1} (-S_{y}/S_{x})$ becomes that in Fig. \ref{nonaxivor}(c), where $\phi(x,y)$ have a branch cut at $y=0$ for $-W(4 \alpha \lambda^{2})/\sqrt{\alpha} \leq x \leq +W(4 \alpha \lambda^{2})/\sqrt{\alpha}$.
\end{thebibliography}
\end{document}